\documentclass[fleqn,10pt]{wlscirep}
\title{Reversible tuning of magnetocaloric Ni-Mn-Ga-Co films on ferroelectric PMN-PT substrates}

\author[1,2,*]{Benjamin Schleicher}
\author[1]{Robert Niemann}
\author[1]{Stefan Schwabe}
\author[1]{Ruben H\"uhne}
\author[1,2]{Ludwig Schultz}
\author[1,2]{Kornelius Nielsch}
\author[1,2]{Sebastian F\"ahler}
\affil[1]{IFW Dresden, Institute for Metallic Materials, Helmholtzstr. 20, D-01069 Dresden, Germany}
\affil[2]{TU Dresden, Institute of Solid State Physics, D-01062 Dresden, Germany}

\affil[*]{b.schleicher@ifw-dresden.de}

\usepackage{upgreek}

\begin{abstract}
Tuning functional properties of thin caloric films by mechanical stress is currently of high interest. In particular, a controllable magnetisation or transition temperature is desired for improved usability in magnetocaloric devices. Here, we present results of epitaxial magnetocaloric Ni-Mn-Ga-Co thin films on ferroelectric Pb(Mg$_{1/3}$Nb$_{2/3}$)$_{0.72}$Ti$_{0.28}$O$_3$ (PMN-PT) substrates. Utilizing X-ray diffraction measurements, we demonstrate that the strain induced in the substrate by application of an electric field can be transferred to the thin film, resulting in a change of the lattice parameters. We examined the consequences of this strain on the magnetic properties of the thin film by temperature and electric field dependent measurements. We did not observe a change of martensitic transformation temperature but a reversible change of magnetisation within the austenitic state, which we attribute to the intrinsic magnetic instability of this metamagnetic Heusler alloy.
\end{abstract}
\begin{document}

\flushbottom
\maketitle
\thispagestyle{empty}

\section*{Introduction}

Magnetocaloric refrigeration has been proposed as a promising tool for more energy efficient cooling.\cite{Pecharsky1997, Gschneidner2000, GschneidnerJr2005, Oliveira2010, Roy2014, Lyubina2017} One of the materials under investigation is the Heusler alloy Ni-Mn-Ga-Co. This compound exhibits an inverse magnetocaloric effect around room temperature due to a first order phase transition from a ferromagnetic austenite to a weak magnetic martensite with an entropy change of up to 17\,J/(kgK) in a magnetic field of 5\,T.\cite{Fabbrici2011} However, one drawback of a first order transition is the narrow usable temperature range as the transition temperature of the material has to span the entire working region of the cooling device. The transition temperature of the Ni-Mn-based Heusler alloys highly depends on multiple parameters like the composition\cite{Cakir2013, Teichert2015, Gottschall2016}, chemical order\cite{Sanchez-Alarcos2014} and stress.\cite{Heczko2006} However, in many cases it is desired to control the transformation temperature of the final sample when the composition and chemical order cannot be changed anymore. Due to the structural change during the transition from the high to the low temperature phase, it is of high interest how this transition and the magnetic properties of the material can be influenced with mechanical stress.\cite{Faehler2012, Liu2012} Thin films can be easily strained when grown on a piezoelectric single crystal and are used as a model system to study the influence of mechanical stress on the functional properties. Significant attention has recently been given to this strain engineering approach, investigating various materials like ferromagnets,\cite{Tkach2015} semiconductors,\cite{Hui2013} superconductors\cite{Trommler2015} and quantum dots\cite{Zhang2016} on Pb(Mg$_{1/3}$Nb$_{2/3}$)O$_{3}$-PbTiO$_3$ (PMN-PT) substrates and magnetocaloric materials like La$_{0.7}$Ca$_{0.3}$MnO$_{3}$\cite{Moya2012} and FeRh\cite{Cherifi2014, Liu2016} on BaTiO$_{3}$ substrates. Additionally, Ni$_{44}$Co$_{5.2}$Mn$_{36.7}$In$_{14.1}$ ribbons adhered to PMN-PT substrates have been studied.\cite{Gong2015} The results obtained so far on these magnetocaloric materials are promising as the hysteresis accompanying the phase transition was reduced\cite{Moya2012, Cherifi2014, Liu2016} or the transition temperature shifted.\cite{Gong2015} 

Here, we will address the challenge of the relatively large hysteresis of a first order transformation, which often inhibits the reversibility of the phase transition in the low magnetic fields available in practical applications. For this, we will first analyse how the strain is transferred from the substrate to the film and estimate the limits of this approach. Then, we will experimentally probe the suitability to shift the transformation temperature in case of a large hysteresis. Though our analysis will show that and why this approach is not feasible, our experiments indicate for a novel mechanism not requiring a shift of transformation temperature, but utilizing the high sensitivity of magnetic properties on the interatomic distances in the metamagnetic Heusler alloys.

\section*{Results and discussion}

\subsection*{Structural investigation}

First, we analysed the strain transfer from the ferroelectric substrate to the magnetocaloric film. For this, the in-plane and out-of-plane lattice parameters of the PMN-PT and the Ni-Mn-Ga-Co in the austenitic phase were determined with X-ray diffraction measurements (Philips X'Pert, Cu-K$_{\upalpha}$ radiation). Therefore, reciprocal space maps (RSM) of the PMN-PT (013) and (0$\overline{1}$3) as well as Ni-Mn-Ga-Co (026) and (0$\overline{2}$6) reflections were measured for increasing applied electric fields from 0\,kV/cm up to 12\,kV/cm, depending on the temperature. The RSMs were recorded at 300\,K, 320\,K and 355\,K. For the latter two temperatures, the Ni-Mn-Ga-Co was in the austenitic phase. The peaks were fitted with a 2D pseudo-Voigt-like function and the lattice parameters were calculated from the peak positions. An exemplary RSM and the exact fitting function is included in section\,1 of the Supplementary online. Fig.~\ref{fig1} summarises the change of the lattice parameters for the substrate and the thin film in dependence of the applied field for three different temperatures.

\begin{figure}
\includegraphics[width=\linewidth]{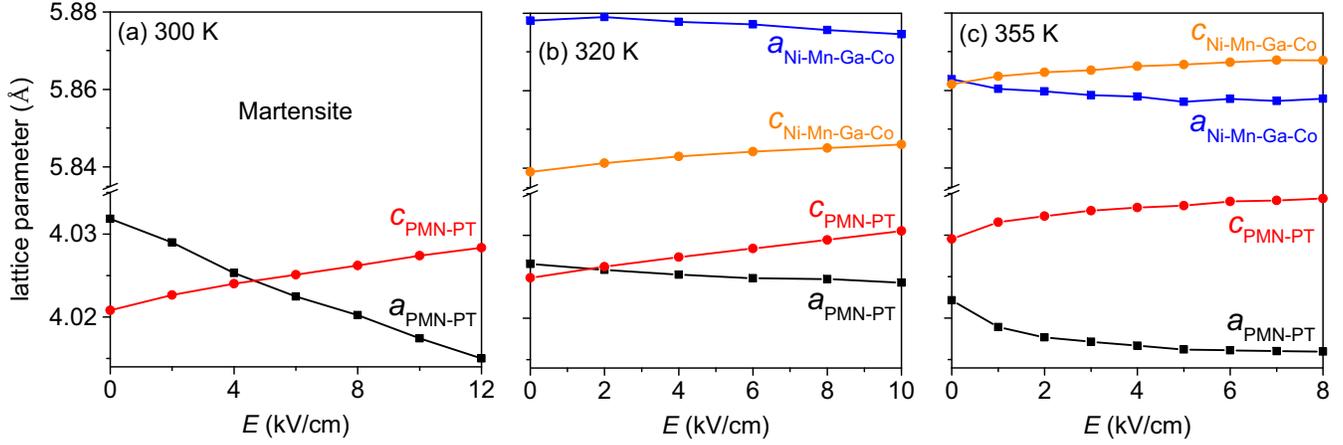}%
\caption{The influence of an applied electric field on the in-plane $a$ and out-of-plane $c$ lattice parameters of the PMN-PT substrate and the austenitic Ni-Mn-Ga-Co film were measured at 300\,K (a), 320\,K (b) and 355\,K (c). Since the Ni-Mn-Ga-Co is martensitic at 300\,K, only the PMN-PT is shown in (a). An in-plane contraction of the substrate and Ni-Mn-Ga-Co (black and blue, respectively) and an elongation in out-of-plane direction (red and orange for PMN-PT and Ni-Mn-Ga-Co, respectively) is visible. \label{fig1}}%
\end{figure}

Below 362\,K, the PMN-PT substrate is in a monoclinic phase,\cite{Herklotz2010} which can be approximated as pseudo-cubic. The same approximation is also possible for the austenitic Ni-Mn-Ga-Co film, exhibiting a small tetragonal distortion between the in-plane and out-of-plane lattice parameters at zero electric field (fig.~\ref{fig1} (b,c)). The difference between the in-plane lattice parameters of substrate and thin film causes a growth with a 45$^\circ$ rotation of their unit cells with respect to each other, which was already shown earlier by pole figure measurements.\cite{Schleicher2015} An increasing electric field leads to a compression of the lattice in-plane ($a$-axis) and an expansion in the out-of-plane direction ($c$-axis). This opposing trend is expected for the conservation of volume and in accordance with previous investigations, where PMN-PT substrates were used for different thin film systems.\cite{Biegalski2010, Kim2013} At 300\,K (fig.~\ref{fig1}(a)), there is the highest influence of the electric field on the PMN-PT lattice parameters with a linear in-plane compression of $\Delta a_{\text{PMN-PT}} = (a_{\text{PMN-PT}}( 12\,\text{kV/cm})-a_{\text{PMN-PT}}(0\,\text{kV/cm}))/ a_{\text{PMN-PT}}(0\,\text{kV/cm}) = -0.46\,\%$ and a linear out-of-plane elongation of $\Delta c_{\text{PMN-PT}} = +0.18\,\%$ for electric field changes of 12\,kV/cm. However, at this temperature the Ni-Mn-Ga-Co is in the martensitic state. Therefore, we cannot give lattice parameters of Ni-Mn-Ga-Co in a pseudo-cubic approximation. In the non-modulated martensite, the spontaneous strain ($c$/$a$-ratio) typically is about 22\,\%,\cite{Niemann2017} which by far exceeds the maximum strain achieved by an electric field. 

At 320\,K, the electric field induced in-plane strain is much smaller than at 300\,K, but completely transferred from the substrate to the thin film with $\Delta a_{\text{PMN-PT}} = \Delta a_{\text{Ni-Mn-Ga-Co}} = -0.06\,\%$ ($\Delta E = 10$\,kV/cm), and the Ni-Mn-Ga-Co is in the austenitic state. This direct coupling within the film plane is expected for a thin film on a thick substrate. As both in-plane directions are assumed to be equivalent, there is a biaxial compressive stress within the magnetocaloric film. For the perpendicular direction, the change of lattice constants is very similar for Ni-Mn-Ga-Co and PMN-PT with $\Delta c_{\text{Ni-Mn-Ga-Co}} = +0.12\,\% $ and $\Delta c_{\text{PMN-PT}} = +0.14\,\%$. As in this direction the film is free to change its thickness, we estimated the Poisson's ratio from $\Delta a_{\text{Ni-Mn-Ga-Co}}$ and $\Delta c_{\text{Ni-Mn-Ga-Co}}$\cite{Biegalski2010} to be
\begin{eqnarray}
\centering
\nu = \frac{\Delta c_{\text{Ni-Mn-Ga-Co}}}{\Delta c_{\text{Ni-Mn-Ga-Co}} - 2 \Delta a_{\text{Ni-Mn-Ga-Co}}} = 0.5 \ . 
\label{eq:poisson}
\end{eqnarray}
This is higher than the literature values of Ni$_2$MnGa in the austenitic phase $\nu = 1/3$,\cite{Kart2010} but a deviation between a bulk sample and a thin film on a substrate can be expected.

To estimate the expected shift of transformation temperature caused by the strain, we compare the biaxial compressive stress in these films with uniaxial tensile stress, as this results in the same deformation of the unit cells. The stress $\sigma$ in the sample, for the maximum strain $\epsilon = -0.06$\,\%, can be calculated from the X-ray diffraction data using the relation\cite{Birkholz2006}
\begin{equation}
\centering
\sigma=\frac{\epsilon \cdot E}{(1+\nu)\sin^2 (\Delta \omega)-2\nu}=7.06\,\text{MPa}
\label{eq:sigma}
\end{equation}
with the Poisson's ratio $\nu = 0.5$ and the offset-angle $\Delta \omega=-18.465^\circ$ of the (026)-reflection. For this estimation, we used $E=10$\,GPa measured on Ni-Mn-Ga single crystals\cite{Chernenko2016} as no values of Co alloyed samples are available. When using the Clausius-Clapeyron equation for the application of mechanical tensile stress along the Ni-Mn-Ga [001] direction\cite{Chernenko2016}
\begin{equation}
\centering
\frac{\text{d}\sigma}{\text{d}T}= 6.0\, \frac{\text{MPa}}{\text{K}} \ ,
\label{eq:CC}
\end{equation}
we obtain a maximum change of transition temperature $\Delta T = +1.2$\,K.

At 355\,K, we observe non-linearities in the strain/electric field curves, untypical for this low field regime (compare, e.\,g. ref\cite{Biegalski2010}). This can be attributed to the vicinity to the phase transition in the PMN-PT from a monoclinic to a tetragonal phase in this electric field-temperature region.\cite{Herklotz2010} We observed that $c > a $ for PMN-PT, which is a consequence of polarization. For the Ni-Mn-Ga-Co film, however, within the measurement accuracy both lattice parameters are equal at very low electric fields. This indicates an almost stress free state at 0\,kV/cm. Presumably, poling of the substrate compensated the thermal and epitaxial strain commonly occurring also in epitaxial Heusler films.\cite{Doyle2008, Thomas2008} 

When comparing $\Delta a_{\text{PMN-PT}}$ for the three different temperatures in fig.~\ref{fig1}, the highest influence of the electric field is at 300\,K ($\Delta a_{\text{PMN-PT}} = -0.42$\,\%). Due to the higher temperature and the reduced maximum electric field, at 320\,K $\Delta a_{\text{PMN-PT}} = -0.06$\,\% is much smaller. Due to the phase transition in the substrate, the $\Delta a_{\text{PMN-PT}} = -0.15$\,\% obtained at 355\,K (for $E=8$\,kV/cm) is larger than at 320\,K. In our experiments, at relatively high temperatures, we were limited to 8\,kV/cm as we observed irreversible shortcuts within the substrate in identically prepared samples when going to higher fields. We attribute this limit to the growth conditions required to form a metallic epitaxial film. Heating the PMN-PT-substrate under UHV conditions presumably results in some evaporation of Pb and loss of oxygen, which degrades the ferroelectric properties.

\subsection*{Magnetic investigation}
 
\begin{figure}
\includegraphics[width=\linewidth]{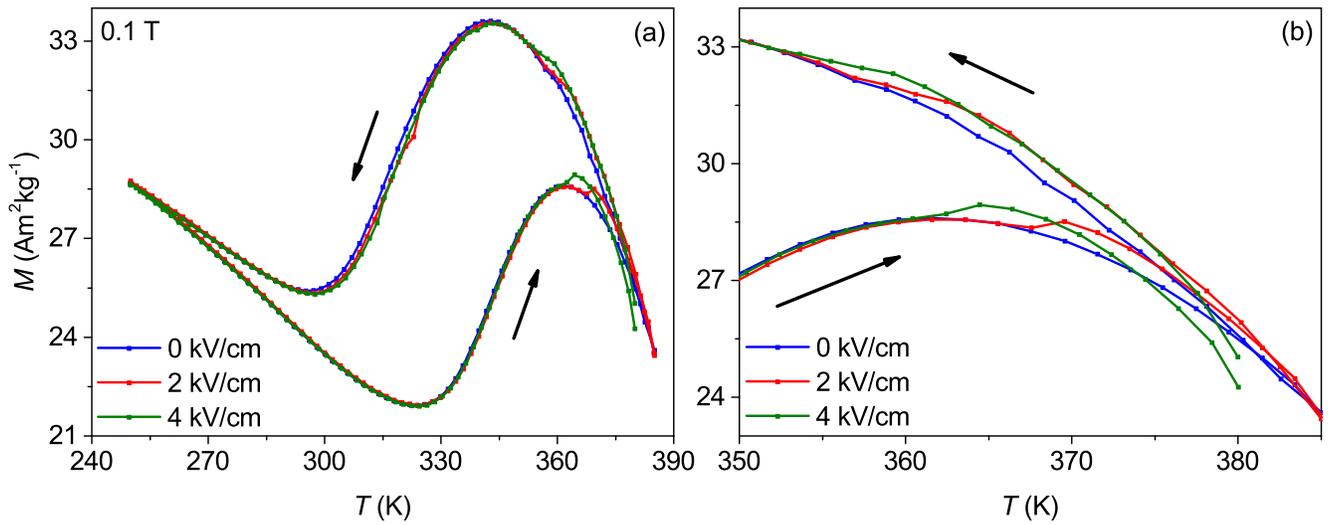}%
\caption{(a) $M$($T$) measurements for $250 \leq T \leq 385$\,K using different electric fields at $\upmu_0 H=0.1$\,T. There is no clear change of the transition temperature visible, but in the austenite region (shown in detail in (b)), a direct influence of the electric field on the magnetisation was observed. \label{fig2}}%
\end{figure}

For the magnetic characterization, a SQUID setup (Quantum Design MPMS) with a custom built sample holder that provided electrical contacts was used. Fig.~\ref{fig2} shows $M$($T$) measurements at a magnetic field of $\upmu_0 H = 0.1$\,T applied in-plane with \mbox{$E$\,=\,0, 2 and 4\,kV/cm} and a temperature sweep rate of 3\,K/min. The starting point of every $M$($T$) loop was at high temperatures to ensure a fully austenitic state in the beginning. The $M$($T$)-measurements at 0 and 2\,kV/cm started at 385\,K. For 4\,kV/cm, the starting temperature was reduced to 380\,K due to an increased risk of an electric breakthrough in the substrate at this higher electric field. For all potential values, we observe the common magnetisation changes of a metamagnetic Heusler alloy.\cite{Nayak2010, Diestel2015} During cooling magnetisation first increases as expected below the Curie temperature of the austenitic phase. Below 340\,K, magnetisation decreases due to the transition to a weak ferromagnetic martensite. Below 300\,K, magnetisation increases again as some ferrimagnetic order within the martensite arises. During heating, the same trends are observed, but the hysteresis of the first order martensitic transition shifts the increase of magnetisation to substantially higher temperatures. In the austenite region, measurements under applied electric field show the appearance of steps in the magnetisation in the cooling as well as heating branch (see enlargement, fig.~\ref{fig2}(b)). In the martensite region, there is no influence of the electric field at all. We do not observe a clear shift of the transformation temperature but some scattering of about $\pm 1$\,K, which we take as resolution limit of this setup. This, however, is higher than the maximum expected shift of $+0.5$\,K from the Clausius-Clapeyron equation~(\ref{eq:CC}) when considering that the maximum electric field in the $M$($T$)-measurements was 4\,kV/cm. It is interesting to put this $\Delta T$ in relation to the observed hysteresis of 24\,K. Thus, the usable potential range is not large enough to overcome hysteresis in these epitaxial films. 

\begin{figure}
\includegraphics[width=\linewidth]{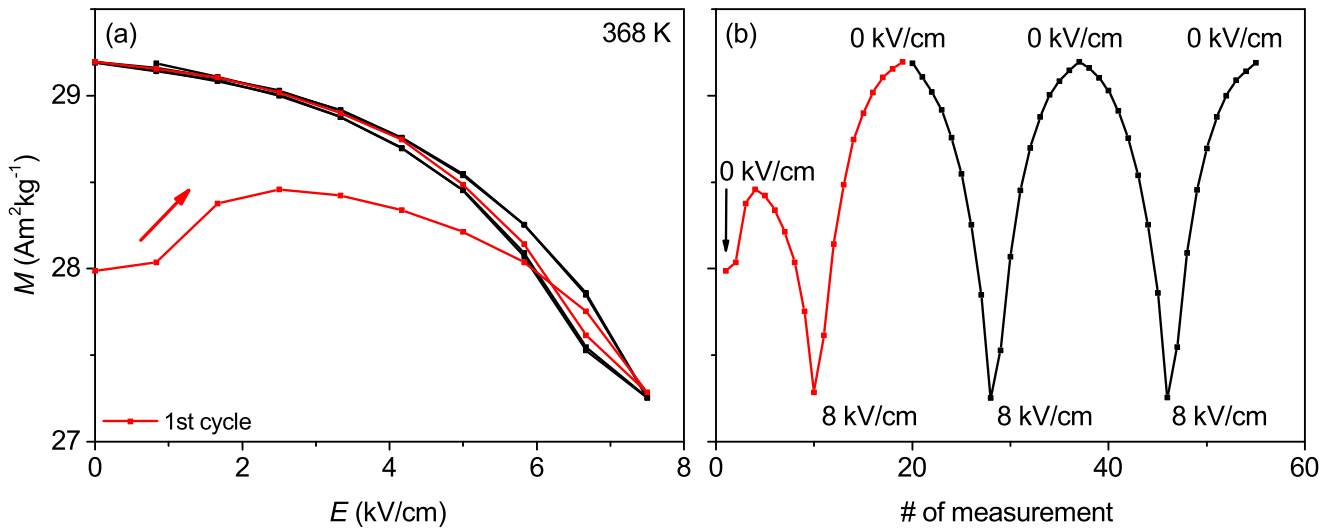}%
\caption{Change of magnetisation $M$ for three sweeps of the electric field with $0 \leq E \leq 8$\,kV/cm at 368\,K and $\upmu_0 H=0.1$\,T. The first sweep is marked red and shows a virgin effect, whereas all sequential field sweeps lead to a reversible change of the magnetisation. (a) shows the $M$($E$)-data, whereas in (b) $M$ is plotted in dependency of the measurement sequence for better visibility of the reproducibility. \label{fig3}}%
\end{figure}

To understand the unexpected change of magnetisation in temperature loops under electric fields especially within the austenite (fig.~\ref{fig2}(b)), $M$($E$) measurements were performed at different fixed temperatures in the austenite, martensite and phase transition region. Fig.~\ref{fig3} shows the resulting changes of the magnetisation for $0 \leq E \leq 8$\,kV/cm in the austenitic phase at 368\,K. Prior to this measurement, the sample was cooled to 250\,K to ensure being on the heating branch of the $M$($T$)-loop seen in fig.~\ref{fig2}.

After a virgin effect during the first electric field sweep (shown in red in fig.~\ref{fig3}), where $M$ first increases and then decreases, a completely reversible change of the magnetisation with $\Delta M \approx 7$\,\% is observed. We attribute the virgin effect to the fact that these experiments are performed in a state, where still some martensite exists in the predominantly austenitic sample. This metastable state is evident from the still open hysteresis at 368\,K in fig.~\ref{fig2}(b). The reversible behaviour for all following cycles must have a different origin, as the maximum potential applicable is not sufficient to shift the transformation temperatures. As an alternative mechanism, we propose to consider the elastic straining of the lattice constants of the austenite (as observed in fig.~\ref{fig1}). Following the analysis of the XRD measurements, an electric field of 8\,kV/cm changes the tetragonality of the Ni-Mn-Ga-Co unit cell by around 0.24\,\%. Though this is low compared to the spontaneous strain of about 22\,\% occurring during the martensitic transition from austenite to non-modulated martensite, the same underlying mechanics may be relevant. For a complete metamagnetic transition, this spontaneous strain changes magnetisation by 100\,\%, which is commonly attributed to the Mn atoms being at an interatomic distance close to the crossover from ferromagnetic to antiferromagnetic coupling.\cite{Buchelnikov2008, Comtesse2014, Priolkar2014} We propose that the high sensitivity of magnetism on the interatomic distances is also the underlying mechanism for the observed reversible change of magnetisation by 7\,\% at a biaxial in-plane compression of about 0.1\,\%. The strain-induced change of the magnetic coupling might also influence the Curie-temperature, but the limitation of the SQUID-device to $T \leq 395$\,K and the high risk of an electric breakthrough at such high temperatures hinders the direct measurement of $T_\text{C}$. Further density functional theory calculations, which are beyond the scope of this experimental paper, are required to confirm, if these slight changes of lattice parameters are sufficient to explain the substantial change of magnetisation.

\begin{figure}
\includegraphics[width=\linewidth]{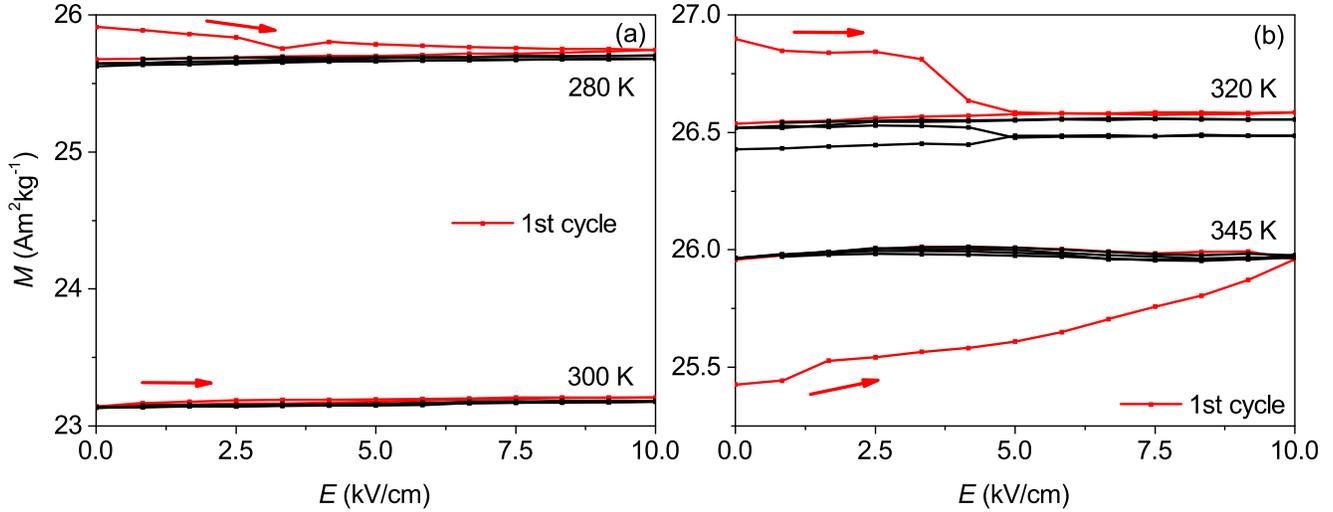}%
\caption{Dependence of the magnetisation within the martensitic and mixed state on the electric field $E$ applied to the PMN-PT substrate. (a) $M$($E$) on heating (300\,K) and cooling branch (280\,K) in the martensitic phase. The magnetisation is not changed reversibly. (b) $M$($E$) in the phase transition region on the cooling (320\,K) and heating (345\,K) branch. After an initial change of the magnetisation during the first cycle, the electric field change has almost no influence on the magnetisation anymore. All measurements were performed at $\upmu_0 H=0.1$\,T. \label{fig4}}%
\end{figure}

Similar $M$($E$) measurements were also performed at temperatures corresponding to different states of the sample. Fig.~\ref{fig4}(a) shows the $M$($E$) measurements in the martensitic phase at 280\,K on the cooling branch and 300\,K on the heating branch. The electric field dependence of the magnetisation in the phase transition region for the transformation from austenite to martensite and vice versa at 320\,K and 345\,K, respectively, is shown in fig.~\ref{fig4}(b). The maximum electric field was increased to 10\,kV/cm as the breakthrough voltage increases with reduced temperature. Again, the first cycle is shown in red.

In the martensitic state, no reversible change of $M$ was observed and the $\Delta M$ almost vanished to a value below 0.1\,\% (see fig.~\ref{fig4}(a)). We attribute this to the ability of the martensite to compensate the mechanical stress by a reorientation of the martensitic variants due to the highly mobile twin boundaries.\cite{Thomas2009, Ranzieri2015} Therefore, the magnetisation in the low temperature phase is not changed. Within the transition region, a slight influence of the mechanical stress on the magnetisation is observed (fig.~\ref{fig4}(b)). In this case, an initial change of the magnetisation in the order of 1\,\% was observed in the cooling (320\,K) as well as the heating (345\,K) branch of the magnetisation loop, when the electric field was applied for the first time. In all following cycles, no reversible change was observed. Noticeable is the different sign of this initial magnetisation change in both measurements. During the first cycle, the magnetisation increases in the heating branch while it decreases in the cooling branch. In other words, the mechanical stress narrows the hysteresis gap between the cooling and heating branch. This indicates that mechanical stress in this metastable coexistence region results in variant reorientation, which follows isothermal kinetics\cite{Perez-Landazabal2012} towards the equilibrium state being in between both branches. All these measurements have been performed in a low magnetic field of 0.1\,T, which may not be sufficient to saturate the sample completely in case of a high anisotropy. To exclude the role of magnetostriction, we performed additional experiments in a sufficiently high magnetic field of 2\,T (see Supplementary Fig.\,S2 and S3 online) resulting in the same behaviour.

\section*{Summery}

To conclude, we investigated the influence of strain created by a ferroelectric substrate on the martensitic phase transition and magnetic properties of a Ni-Mn-Ga-Co thin film. It was demonstrated that the in-plane and out-of-plane lattice parameters of the substrate and austenitic thin film can indeed be tuned by the application of an electric field to the multiferroic stack. Although a relatively large reversible change of the magnetisation in the austenite phase was observed, the transformation temperature could not be varied. This is in contradiction to reports on Ni-Mn-In-Co ribbons adhered to PMN-PT, where a shift of the transition temperature for different applied electric fields was observed.\cite{Gong2015} We attribute this to differences in the microstructure. In polycrystalline Ni-Mn-In-Co ribbons, the compensation of mechanical stress by twin boundary movement in the martensite is significantly limited due to grain boundaries.\cite{Romberg2013} In contrast, our single-crystalline films allow for an easy compensation of stress since twin boundary movement is possible. We attribute the reversible change of magnetisation, only observed within the austenitic state, to a different mechanism. We suggest that it originates from the change of interatomic distances, which have a strong influence on the ferromagnetic-antiferromagnetic coupling between Mn atoms. The advantage of this mechanism is its reversibility, avoiding the large hysteresis of the first order phase transition. Though the change of the magnetisation is significantly lower compared to the one of the martensitic transition, utilizing the intrinsic magnetic instability of metamagnetic Heusler alloys may become an interesting approach for a reversible tuning of magnetocaloric properties.

\section*{Methods}

As described previously,\cite{Schleicher2015} epitaxial Ni-Mn-Ga-Co thin films were grown by DC magnetron sputter deposition on 0.3\,mm thick single-crystalline PMN-PT substrates provided by Morgan Electroceramics. The Ni$_{43}$Mn$_{32}$Ga$_{20}$Co$_{5}$ films are 350\,nm thick and a 20\,nm Cr buffer layer was used to enhance epitaxial growth. The epitaxial relationship obtained by XRD-measurements is PMN-PT(001)[110]$\parallel$Cr(001)[100]$\parallel$Ni-Mn-Ga-Co(001)[100]. For the measurements with an applied electric voltage $U$, a Au/NiCr layer on the unpolished side of the substrate was used as bottom electrode whereas the conducting thin film was used as top electrode. More details on the sample preparation as well as first structural and magnetic characterizations can be found elsewhere.\cite{Schleicher2015} For the stress-dependent measurements, a Keithley sourcemeter was used to apply an electric voltage to the multiferroic stack. In order to protect the samples from voltage breakthrough, the current was limited to 105\,$\upmu$A and the maximum voltage applied was 300\,V. For the magnetisation measurements in dependence of temperature $T$ and voltage $U$, the temperature sweep rate was 3\,K/min and the voltage step size used was 25\,V. Since the electric field is the more physical and material(-thickness) independent parameter, the applied voltage was converted to the electric field $E$ and for better readability rounded to full digits. Before every measurement series with applied electric field, the substrate was polarized along the [001] direction to eliminate the influence of the hysteretic behaviour of the PMN-PT substrates.\cite{Yang2014} As we experienced cracking of one sample during demounting, we had to use two samples for structural analysis and another one for the magnetic characterization. All samples have been prepared under identical conditions and have the same composition within the measurement accuracy of about 1\,at.\% but differ about 40\,K in the transition temperature. However, it is well known for Ni-Mn-based alloys that the transition temperature can drastically change even with small differences in composition\cite{Gottschall2016a} whereas all other properties are nonetheless practically equal.

\textbf{Data availability:} The datasets generated and analysed during the current study are available from the corresponding author on reasonable request.


\section*{Acknowledgements}

Funding by DFG through SPP 1599 www.FerroicCooling.de grant no FA 453/11 and HU 1726/3 is gratefully acknowledged.

\section*{Author contributions statement}

B.S., R.N. and S.F. conceived the experiments. B.S. conducted all experiments and analysed the SQUID measurements. R.N., S.S. and R.H. analysed the RSM measurements. B.S., S.F. and R.H. interpreted the results. L.S. and K.N. supervised the work of B.S. and S.S. All authors reviewed the manuscript. 

\section*{Additional information}

\textbf{Competing financial interests:} The authors declare no competing financial interests.

\end{document}